\documentclass[twocolumn,pre,epsfig,rotate,noshowpacs,cjk,superscriptaddress,english]{revtex4}
\usepackage{graphicx}
\usepackage{amsmath}
\usepackage{dcolumn}
\usepackage{epsfig}
\usepackage{bm}
\begin{document}

\title{Crossover from ballistic to normal heat transport in the $\phi^{4}$ lattice:
\\ If nonconservation of momentum is the reason, what is the mechanism?}

\author{Daxing Xiong}
\email{phyxiongdx@fzu.edu.cn}
\affiliation{Department of Physics,
Fuzhou University, Fuzhou 350108, Fujian, China}

\author{Danial Saadatmand}
\email{saadatmand.d@gmail.com}
\affiliation{Department of Physics,
University of Sistan and Baluchestan, Zahedan, Iran}

\author{Sergey V. Dmitriev}
\email{dmitriev.sergey.v@gmail.com} \affiliation{Institute for
Metals Superplasticity Problems of RAS, Khalturin St. 39, 450001 Ufa,
Russia}
\affiliation{National Research Tomsk State University, Lenin
Avenue 36, 634050 Tomsk, Russia}

\begin{abstract}
Anomalous (non-Fourier's) heat transport is no longer just a
theoretical issue since it has been observed experimentally in a
number of low-dimensional nanomaterials, such as SiGe nanowires,
carbon nanotubes, and others. To understand these anomalous
behaviors, exploring the microscopic origin of normal (Fourier's)
heat transport is a fascinating theoretical topic. However, this
issue has not yet been fully understood even for one-dimensional
(1D) model chains, in spite of a great amount of thorough studies
done to date. From those studies it has been widely accepted that
the conservation of momentum is a key ingredient to induce anomalous
heat transport, while momentum-nonconserving systems usually support
normal heat transport where Fourier's law is valid. But if the
nonconservation of momentum is the reason, what is the underlying
microscopic mechanism for the observed normal heat transport? Here
we carefully revisit a typical 1D momentum-nonconserving $\phi^{4}$
model and present evidence that the mobile discrete breathers or, in
other words, the moving intrinsic localized modes with frequency
components above the linear phonon band can be responsible for that.
\end{abstract}
\maketitle

\section{Introduction}
As one of the vivid examples for studying the microscopic origin of
the macroscopic irreversibility in terms of deterministic dynamics,
heat transport in one-dimensional (1D) lattice systems has attracted
intensive theoretical studies for several
decades~\cite{Lepri_Report,Dhar_Report,2016Book}. A challenging
problem here is to justify the microscopic origin of the Fourier's
law, which states
\begin{equation}
\emph{J}= - \kappa \frac{\partial T} {\partial x},
\end{equation}
where $\emph{J}\,$ is the heat current, $\kappa$ is the material
constant called thermal conductivity, and $\frac{\partial T}
{\partial x}$ is the spatial temperature gradient. However, in a
number of theoretical studies, it has been demonstrated that for
general 1D lattices, $\kappa$ is not just an intrinsic property of a
material but depends on the chain's length $L$ following $\kappa\sim
L^{\alpha}$ with $0 \le \alpha \le 1$. The limiting case of
$\alpha=1$ corresponds to the ballistic thermal transport and for
$\alpha=0$ one has normal heat conduction obeying Fourier's law,
whereas the cases of $0<\alpha <1$ are usually called the
superdiffusive heat transport.

With the development of laser technology~\cite{Laser1,Laser2} which
allows to probe the thermal properties of materials at reduced size
and time scales, ballistic thermal conduction ($\alpha=1$) has been
observed experimentally for Si$_{0.9}$Ge$_{0.1}$ and
Si$_{0.4}$Ge$_{0.6}$ nanowires, carbon nanotubes, holey silicon,
Al$_{0.1}$Ga$_{0.9}$N thin film, and
others~\cite{2016BookChang,2017LWLLC,2015HHCLCLC,2013HCLCLC,holeySi,AlGaN}.
In those studies, this ballistic phonon transport has been found for
(quasi)-1D samples having length less than a threshold value, $L^*$
(usually characterized by the phonon mean free path), such as that
for SiGe nanowires $L^* \approx 8.3\,\mu$m \cite{2013HCLCLC}, for
carbon nanotubes $L^* \approx 1$~mm \cite{2017LWLLC}, and for holey
silicon $L^* \approx 200$~nm \cite{holeySi}. Thus, these studies
suggest that the ballistic thermal conduction in real (quasi-)1D
materials can persist over macroscopic distances. In fact, it has
been usually believed that in such materials heat is conducted
ballistically by the low-frequency, long-wavelength
phonons~\cite{2016BookChang}, and hence the anomalous ballistic
behavior can persist even with the presence of defects, isotopic
disorders, impurities, and surface absorbates~\cite{2017LWLLC}.

Beyond ballistic transport, the superdiffusive conduction ($0<\alpha
<1$), can be found when the samples' length and time scales are
comparable to the phonon mean free path, in particular in some
nanoparticle embedded semiconductor alloys~\cite{Ver3,Upadhyaya}.
Although this is just a transient process and finally the normal
diffusive behavior ($\alpha=1$) will appear as the length and time
scales increase further, this superdiffusive heat transport is very
peculiar and can be understood by a truncated L\'{e}vy
formalism~\cite{Ver3,Upadhyaya,Ver1,Ver2}, thus suggesting a fractal
L\'{e}vy heat transport physics rather than the usual Brownian
motion.

To further interpret this transition from ballistic to normal
behavior, a recent proposed kinetic-collective model~\cite{Coll1}
based on different phonon-phonon scattering physics might be
worthwhile, from which a wide range of temperature- and
size-dependent thermal conductivity can be predicted quite
well~\cite{Coll1,Coll2,Coll3}. However, obviously such a theoretical
model does not involve the effects of other nonlinear excitations,
such as solitons~\cite{soliton} and discrete breathers
(DBs)~\cite{DBsReview-1, DBsReview-2, DBsReview-3, DBsReview-4}.

In addition to the experimental investigations, there are also a
number of molecular dynamics studies on quasi-1D and 2D
nanomaterials \cite{Suppression,SavinCNT,Isotop,WdW}. Such studies
showed that the rough edges of long graphene nanoribbons suppress
thermal conductivity by two orders of magnitude~\cite{Suppression}.
Isolated carbon nanotube demonstrates anomalous heat transport due
to the long-wavelength acoustic phonons, while nanotube interacting
with a flat substrate displays normal thermal conductivity due to
both the appearance of a gap in the acoustic phonon spectrum and the
absorption of long-wavelength acoustic phonons by the
substrate~\cite{SavinCNT}. Isotopic doping decreases thermal
conductivity of silicene nanosheets~\cite{Isotop}.

Coming back to 1D model chains, up to now, many ingredients related
to anomalous thermal conduction have been carefully considered,
among them are
chaos~\cite{Chaos-1,Chaos-2,Chaos-3,Chaos-4,Chaos-5,Chaos-6},
conservation of
momentum~\cite{Momentum-1,Momentum-2,Momentum-3,Momentum-4},
asymmetric interactions~\cite{Asy-1,Asy-2,Asy-3,Asy-4,Asy-5},
linearity \cite{Chaos-6,Rieder,Krivtsov1,Krivtsov2,Krivtsov3},
integrability~\cite{Benenti}, pressure~\cite{Pressure-1,Pressure-2},
surface scattering effect~\cite{SSE} etc. Particularly, the
viewpoint that a system with (without) a momentum conservation
property should disobey (obey) Fourier's law has been widely
accepted, although there are still some contradictory results on
this
issue~\cite{Contradictory-1,Contradictory-2,Contradictory-3,Contradictory-4,Contradictory-5,Contradictory-6,Contradictory-7,Xiong2016-1,Xiong2016-2}.
A suggested picture to support this viewpoint is that the on-site
potentials which destroy the conservation of momentum can induce a
strong scattering mechanism for phonons, eventually resulting in the
diffusive Fourier's heat transport. In some
literature~\cite{DBs-1,DBs-2}, such scattering mechanism is usually
attributed to DBs. Nevertheless, at present, our understanding of
the physical picture of this scattering process is still lacking.

In this work we therefore revisit a typical momentum-nonconserving
system, i.e., the $\phi^{4}$ model with a single-well on-site
potential. Keeping in mind that only the nonlinear on-site potential
can result in normal heat transport~\cite{Chaos-6}, we shall
carefully examine both the system's heat transport and DBs
properties under different strengths of the nonlinearity, trying to
establish a connection between them. Our aim is to present a
possible mechanism for demonstration why a system with nonconserved
momentum can generally lead to normal heat transport.

The rest of this article is organized as follows: In
Sec.~\ref{SecModel} we first introduce the reference model,
demonstrate its linear phonon dispersion, and discuss what kind of
DBs are likely to exist for the given type of nonlinearity.
Sec.~\ref{SecMethos} then describes our numerical approaches used
in the heat transport study. In Sec.~\ref{SecResults}, we provide our main
results, from which we will see a close relationship between heat
transport and DBs properties, with the change of the strength of the
nonlinearity. Sec.~\ref{SecDiscussion} presents a discussion and
finally Sec.~\ref{SecConclusions} draws our conclusions.

\section{Model}
\label{SecModel}
\begin{figure}
\begin{centering}
\vspace{-.6cm} \includegraphics[width=8cm]{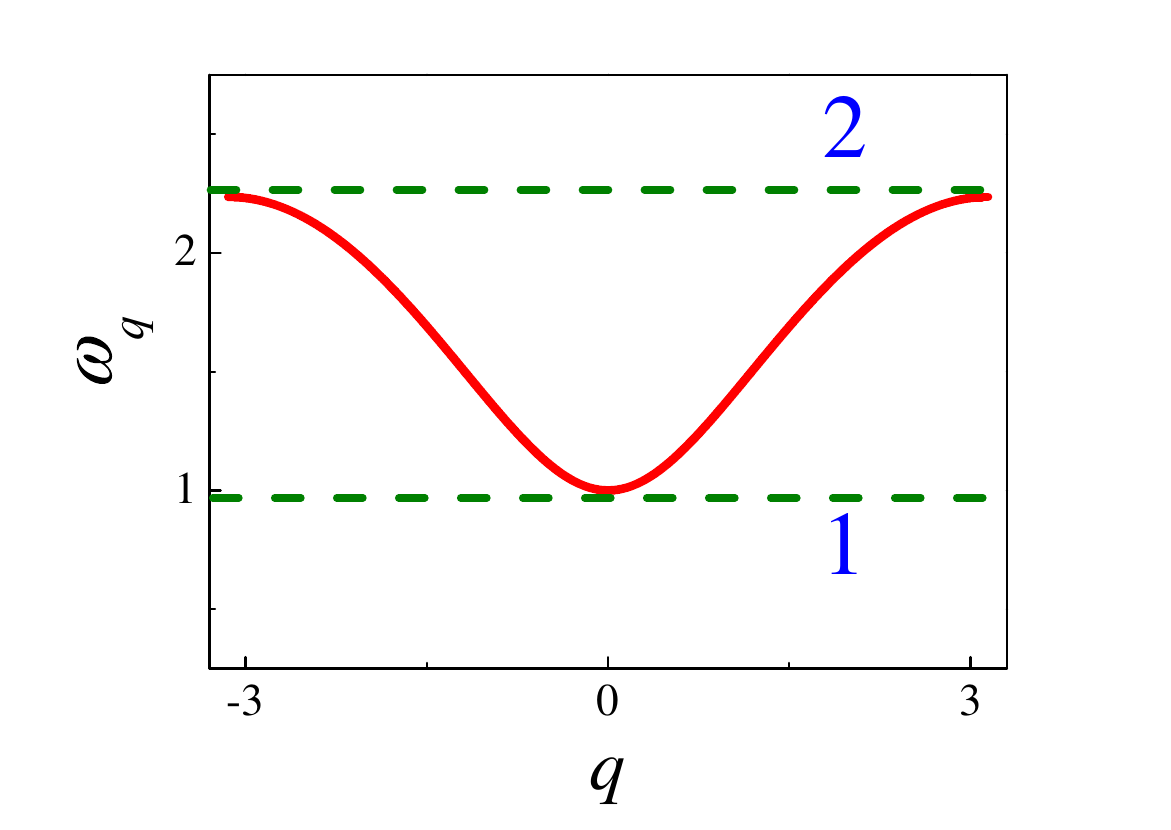} \vspace{-.6cm}
\caption{\label{fig1} The phonon dispersion relation for the
$\phi^{4}$ lattice, where the lines 1 and 2 indicate the lower and
upper edges of the linear phonon band. Note that for the considered
hard-type anharmonicity, only DBs with frequencies above line $2$
can exist.}\vspace{-.3cm}
\end{centering}
\end{figure}
The Hamiltonian of 1D $\phi^{4}$ lattice is given
by
\begin{equation}\label{Ham}
H= \sum_{k=1}^{L} \frac{p_k^2}{2}+V(\phi_{k+1}-\phi_{k})+ U(\phi_k).
\end{equation}
Here $L$ particles are considered; all of them have unit mass; $p_k$
is the momentum of the $k$th particle; $\phi_k$ is its displacement
from equilibrium position; $V= \frac{\xi^2}{2}$ and
$U=\frac{\xi^2}{2}+ \beta \frac{\xi^4}{4}$ denote the harmonic
interparticle interaction and the nonlinear on-site potential,
respectively. Parameter $\beta$ ($\beta>0$) controls the strength of
the nonlinearity in $U$, variation of which will help us to explore
the detailed relationship between the heat transport and the DBs
properties. We note that for $\beta > 0$ the on-site potential focused here is single-well
and of hard-type anharmonicity.

From Eq.~\eqref{Ham}, the following equations of motion can be
derived
\begin{equation}
\frac{\rm{d}^2\phi_{\mit{k}}}{\rm{d} \mit{t}^{\rm{2}}}=
(\phi_{k-1}-2\phi_k+\phi_{k+1}) - \phi_k -\beta \phi_k^3.
\end{equation}
Then, under harmonic approximation ($\beta=0$), the system's phonon
dispersion relation can be easily obtained:
\begin{equation} \label{dispersion}
\omega_q= \sqrt{4 \sin^2 \Big(\frac{q}{2}\Big)+1},
\end{equation}
where $q$ is the wave number and $\omega_q$ is the corresponding
phonon frequency. Such phonon dispersion relation is shown in
Fig.~\ref{fig1}. As can be seen, unlike the linear phonon dispersion
of a momentum-conserving system without including the on-site
potential, the lowest wavelength phonon frequency in our case is now
no longer at the zero value, instead, its frequency is shifted to
the value of $\omega_q^{\min}=1$. On the other hand, the shortest
wavelength phonon frequency is equal to
$\omega_q^{\max}=\sqrt{5}\approx 2.236$. Accordingly, it is easy to
find that the phonon's group velocity $v_g$, defined as
$v_g=\frac{\rm{d} \omega_{\mit{q}}} {\rm{d} \mit{q}}$, vanishes for
$q\rightarrow 0$ and $q\rightarrow \pm\pi$, which is a necessary
condition for the excitation of DBs when including the nonlinearity.

DBs frequency must lie outside the phonon spectrum. For the
hard-type nonlinearity considered ($\beta>0$), only DBs with
frequencies lying in the range denoted by $2$ in Fig.~\ref{fig1},
i.e., above the linear phonon band, are
possible~\cite{DBsReview-1,DBsReview-2,DBsReview-3,DBsReview-4}. It
can also be expected that, the nonlinearity together with the phonon
dispersion as shown in Fig.~\ref{fig1} induced by the on-site
potential will make such kind of DBs properties very peculiar,
different from those in the usually considered momentum-conserving
systems. So, in view of these general understandings, in what
follows we will examine the DBs properties at given strength of the
nonlinearity, aiming to understand their relation to thermal
transport.
\section{Method}
\label{SecMethos} To characterize the system's thermal transport
property, one can employ the equilibrium fluctuation-correlation
method~\cite{Zhao2006,RW-1,RW-2} to derive the spatiotemporal
correlation function of heat fluctuations~\cite{Forster,Liquid}:
\begin{equation} \label{QQ}
\rho_{Q} (m,t)=\frac{\langle \Delta Q_{j}(t) \Delta Q_{i}(0)
\rangle}{\langle \Delta Q_{i}(0) \Delta Q_{i}(0) \rangle}.
\end{equation}
Here $m=j-i$; $\langle \cdot \rangle$ represents the spatiotemporal
average. $Q_i(t)$ is the heat density within a finite volume (bin
$i$) at time $t$, whose expression~\cite{Forster,Liquid}
\begin{equation} \label{Q}
Q_i(t)\equiv E_i(t)-\frac{(\langle E \rangle +\langle F
\rangle)M_i(t)}{\langle M \rangle}
\end{equation}
is obtained from basic thermodynamics in the
textbooks~\cite{Forster,Liquid}. To compute $Q_i(t)$, in practice
one can first divide the 1D lattice into several equivalent bins. In
each bin, we then calculate the number of particles $M_i$, the
energy $E_i$ and the pressure $F_i$ within the bin. Finally, the
heat in the $i$th bin can be derived from~\eqref{Q} and its
fluctuation then is $\Delta Q_i(t)= Q_i(t)-\langle Q_i \rangle$. We
note that since the system is 1D, the pressure is equal to the force
and can be calculated from the gradient of the potential.

This simulation approach for deriving $\rho_{Q} (m,t)$ has been
widely used in many
publications~\cite{Xiong2016-1,Xiong2016-2,Zhao2006,RW-1,RW-2,Chen2013,Xiong2017}.
In particular, from the perspective of random walks
theory~\cite{Zhao2006,RW-1,RW-2,Chen2013}, $\rho_{Q} (m,t)$
(normalized) can be viewed as the heat spreading density, which
together with a space-time scaling analysis has been suggested being
able to provide very detailed information for characterizing the
corresponding thermal transport behavior~\cite{Dhar2014}. In the
hydrodynamics theory, $\rho_{Q} (m,t)$ is believed to correspond to
the heat mode's correlation~\cite{Chen2013,Spohn2014}.

To simulate $\rho_Q(m,t)$, we consider a chain with $L=4001$
particles, which allows an initial heat fluctuation located at the
center to spread out a lag time at least up to $t=1500$. This is
because the introduction of the on-site potential reduces the group
velocity of phonons as compared to the momentum-conserving harmonic
chain (its phonon group velocity is unity). We set both the
equilibrium distance between the particles as well as the lattice
constant to unity. This means that the number of particles $L$ is
equal to the system size. So, for a system with symmetric (even)
type interactions, it can be easily inferred that the average
pressure $\langle F \rangle$ is always zero. We apply periodic
boundary conditions and fix the number of bins to be $(L-1)/2$. We
use the stochastic Langevin heat
baths~\cite{Lepri_Report,Dhar_Report} to thermalize the system and
to prepare a canonical equilibrium state with a fixed temperature
$T=0.5$. The nonlinear parameter $\beta$ is varied over a wide range
from $\beta=0.01$ to $\beta=5$. Under this setup, we employ the
Runge-Kutta algorithm of seventh to eighth order with a time step of
$h=0.05$ to evolve the system. Each canonical equilibrium system is
prepared by evolving the system for a long enough time ($>10^7$ time
units) from properly assigned initial random states. Finally, we use
ensembles of about $8\times10^9$ data points to compute the
correlation function.
\section{Results}
\label{SecResults}
\subsection{Heat transport}
\begin{figure}
\begin{centering}
\vspace{-.6cm} \includegraphics[width=8.8cm]{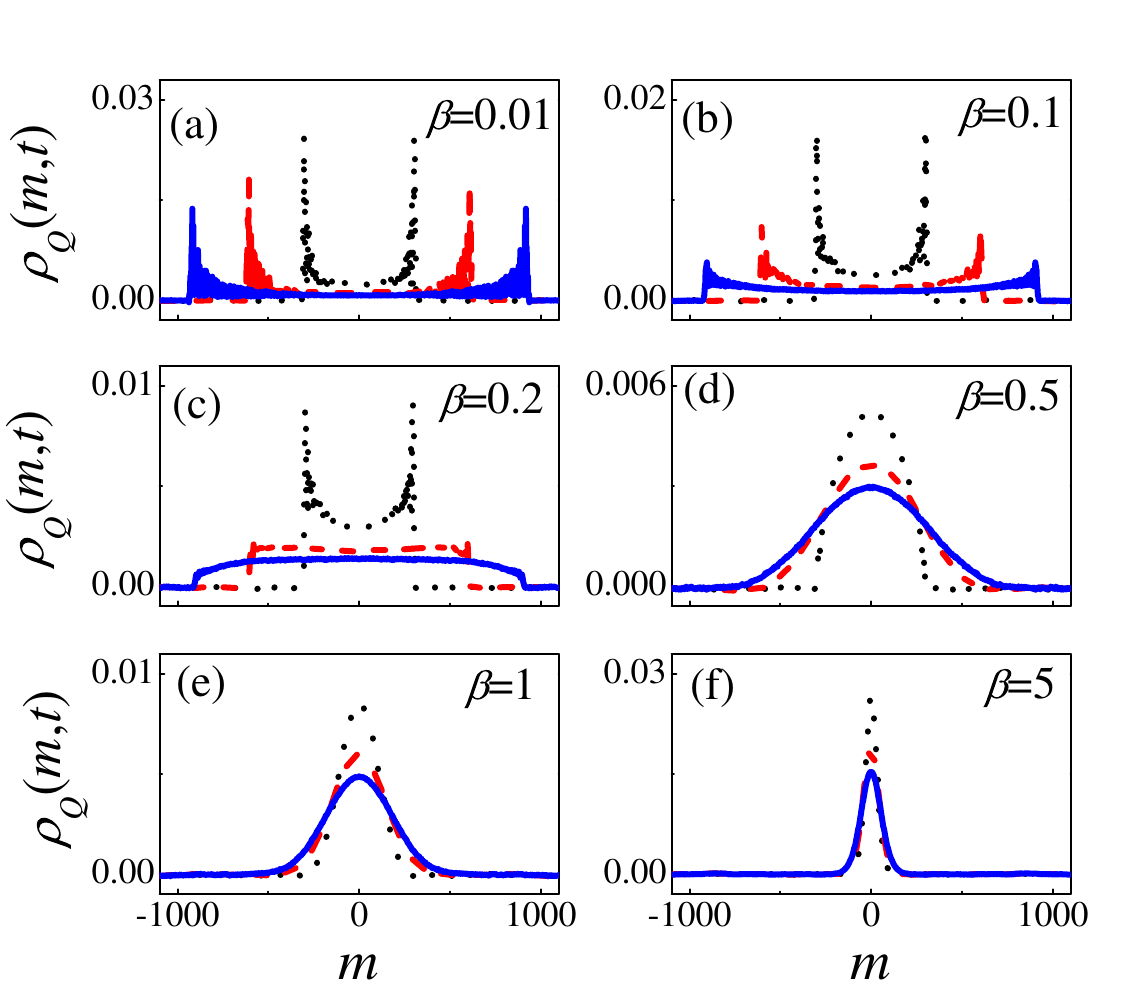} \vspace{-.6cm}
\caption{\label{fig2} $\rho_Q(m,t)$ calculated for several $\beta$ values
at three typical long times $t=500$ (dotted); $t=1000$ (dashed), and
$t=1500$ (solid). }\vspace{-.3cm}
\end{centering}
\end{figure}
We first present several $\beta$-dependent profiles of $\rho_Q(m,t)$
for three long times in Fig.~\ref{fig2}. As can be seen, indeed only
including a strong enough nonlinearity can lead to the perfect Gaussian
profile of $\rho_Q(m,t)$ [see Fig.~\ref{fig2}(f)], which is an
evidence of normal thermal transport obeying Fourier's law. This is
consistent with the argument that the linear optical chain, even
with nonconserved momentum, cannot exhibit normal heat
transport~\cite{Chaos-6}. One can also find that for relatively small
nonlinearity, $\rho_Q(m,t)$ shows a U-shape with some oscillations
[see Fig.~\ref{fig2}(a)]. This is a typical signature of ballistic
heat transport and can be predicted by a recent theory of ``phonon
random walks''~\cite{PRW}. In addition, in the intermediate range of
the nonlinearity, there is a transition (crossover) from ballistic
to normal transport [see Fig.~\ref{fig2}(b)-(e)], i.e., with the
increase of $\beta$, at first the front peaks of $\rho_Q(m,t)$ are
quickly damped, then after this damping process has been almost
finished, the central part of $\rho_Q(m,t)$ starts to be humped and
seems more and more concentrated.
\begin{figure}
\begin{centering}
\vspace{-.6cm} \includegraphics[width=8.8cm]{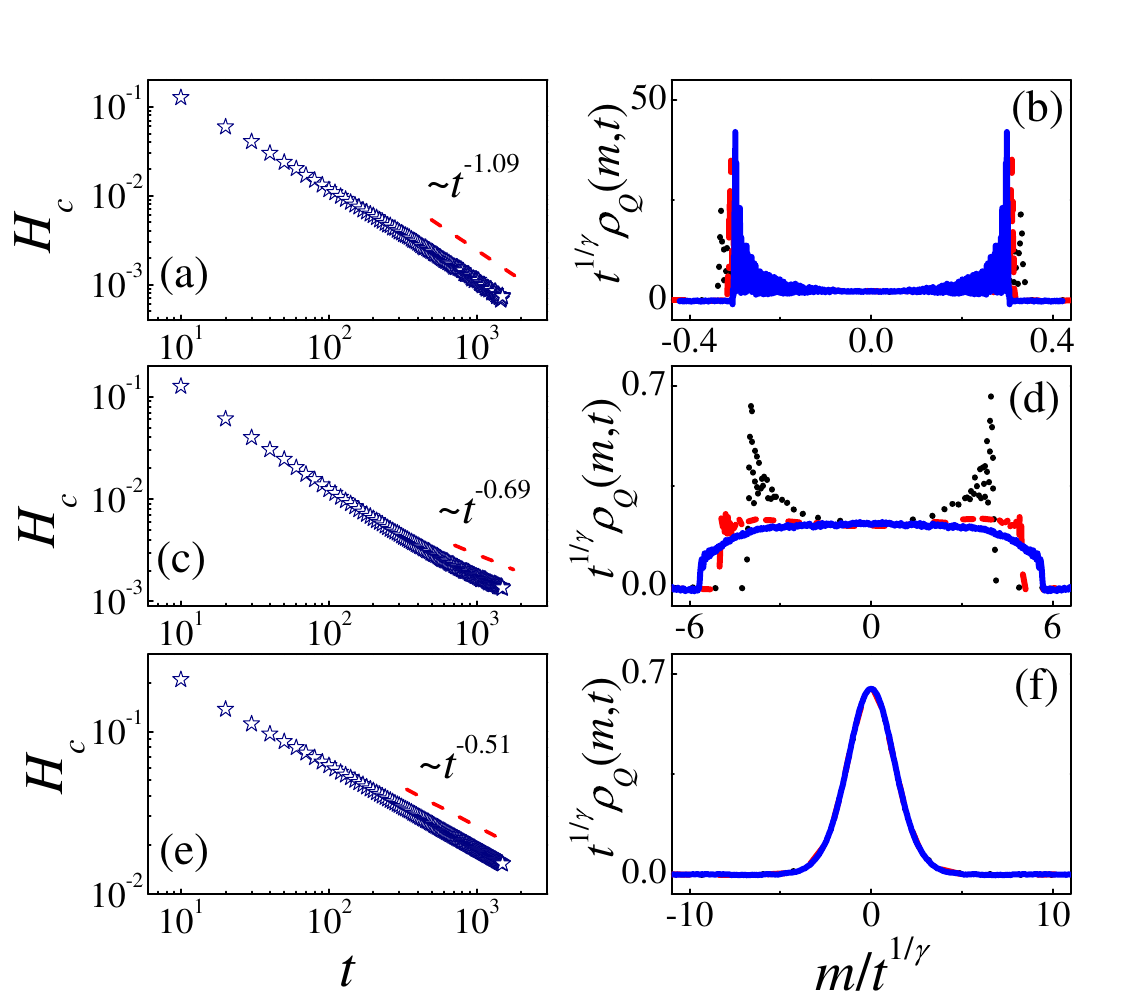} \vspace{-.6cm}
\caption{\label{fig3} Left: (a), (c) and (e), the height $H_c$ of
the central peaks of $\rho_Q(m,t)$ vs $t$ for extracting the scaling
exponent $\gamma$; Right: (b), (d) and (f), the rescaled
$\rho_Q(m,t)$ using formula~\eqref{Scaling}, where (a) and (b) for
$\beta=0.01$; (c) and (d) for $\beta=0.2$; (e) and (f) for
$\beta=5$, respectively. In (b), (d) and (f), three typical long
times of $t=500$ (dotted); $t=1000$ (dashed), and $t=1500$ (solid)
are compared. }\vspace{-.3cm}
\end{centering}
\end{figure}

We are particularly interested in this transition process in the
intermediate range of $\beta$. So, we perform a space-time scaling
analysis~\cite{RW-1,RW-2}
\begin{equation} \label{Scaling}
\rho_Q(m,t) \simeq \frac{1}{t^{1/\gamma}} \rho_Q\left(\frac{m}{
t^{1/\gamma}},t\right)
\end{equation}
of the central part of $\rho_Q(m,t)$ for different $\beta$ values.
Three typical results of the rescaled $\rho_Q(m,t)$ are shown in
Fig.~\ref{fig3}(b)(d)(f). According to the random walk
theory~\cite{RW-1,RW-2}, the scaling exponent $\gamma$ can be
extracted from the time scaling behavior of the height $H_c$ of the
central peaks of $\rho_Q(m,t)$ [see Fig.~\ref{fig3}(a)(c)(e)] and
related to the system size scaling exponent $\alpha$ of heat
conductivity by $\alpha=2-\gamma$~\cite{RW-1,RW-2}. Hence,
$\gamma=1$ and $\gamma=2$ correspond to the ballistic and normal
transport, respectively~\cite{RW-1,RW-2}. So, given these scaling
exponents, Fig.~\ref{fig3}(b) [(f)] indicates the heat transport
close to ballistic (normal); Fig.~\ref{fig3}(d) suggests a crossover
process.
\begin{figure}
\begin{centering}
\vspace{-.6cm} \includegraphics[width=8cm]{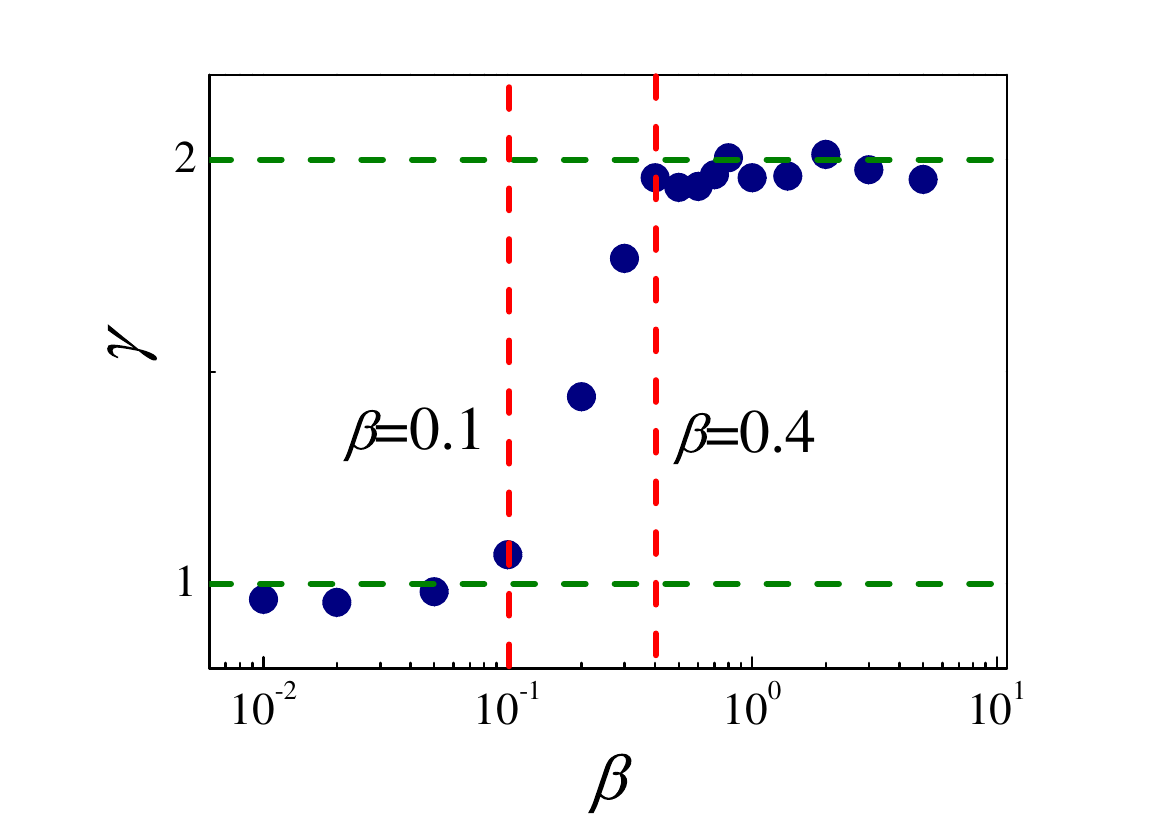} \vspace{-.6cm}
\caption{\label{fig4} The scaling exponent $\gamma$ vs $\beta$ for
indicating the crossover process from ballistic to normal heat
transport, where horizontal lines, from bottom to top, denote
$\gamma=1$ and $\gamma=2$; the vertical lines, from left to right,
represent $\beta=0.1$ and $\beta=0.4$, respectively.}\vspace{-.3cm}
\end{centering}
\end{figure}

To readily capture this crossover process, we also plot the result
of $\gamma$ versus $\beta$ in Fig.~\ref{fig4}. This helps us to more
precisely infer the transition point. As can be seen,
Fig.~\ref{fig4} suggests that the transition point is in between
$\beta=0.1$ and $\beta=0.4$. As to this point, we note that
$\beta=0.2$ is just a central point among them, which is consistent
with the result in Fig.~\ref{fig2}(c), where the front peaks of
$\rho_Q(m,t)$ are nearly completely damped for a relatively long
time. In the very long time and with very long size simulation, one
may thus expect that $\beta_c=0.2$ is probably a transition point
for the system's heat transport crossover from ballistic to normal
at the focused temperature $T=0.5$.
%
%
\begin{figure}
\begin{centering}
\vspace{-.6cm} \includegraphics[width=8.8cm]{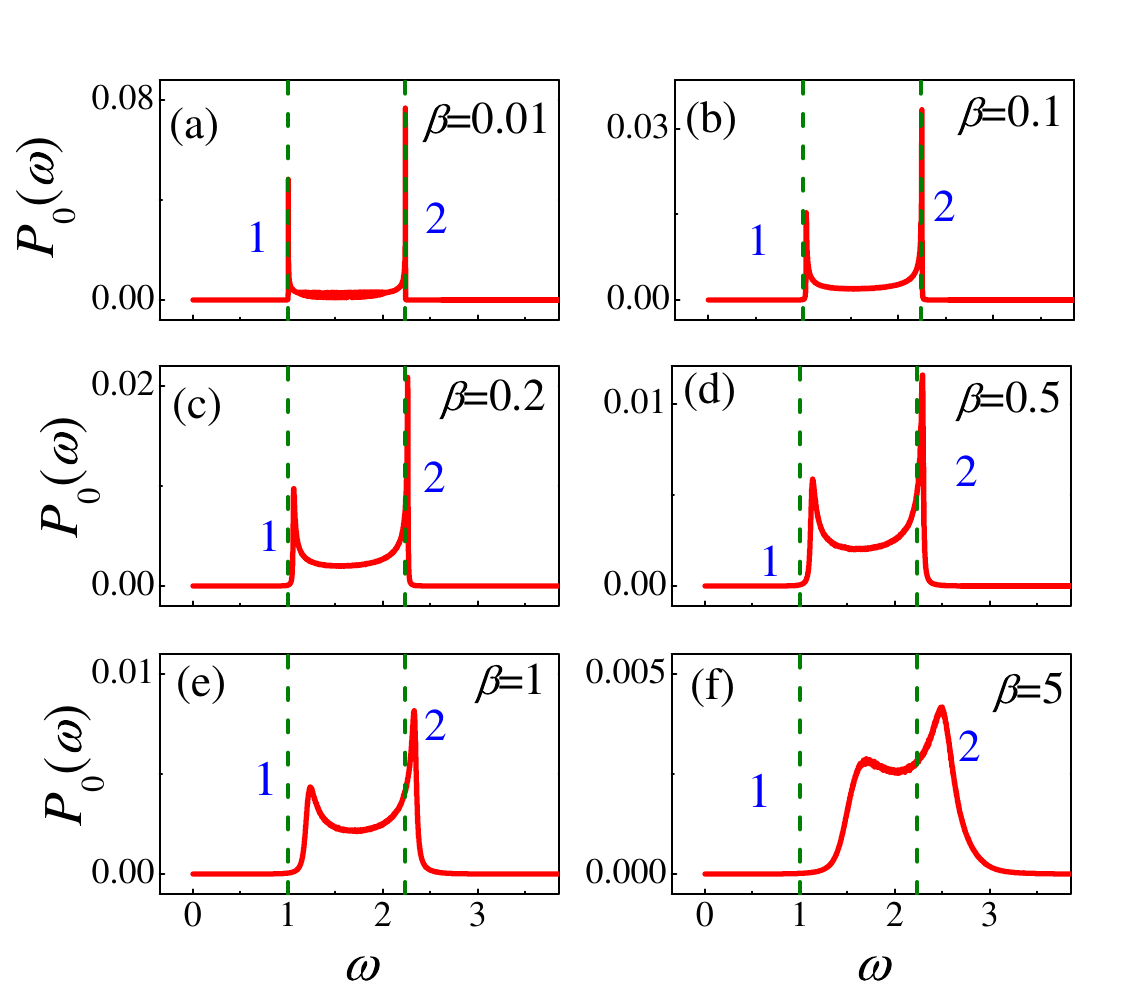} \vspace{-.6cm}
\caption{\label{fig6} The power spectrum $P_0(\omega)$ of the
equilibrium states at $T=0.5$ for the chains with different
strengths of the nonlinearity: (a) $\beta=0.01$, (b) $\beta=0.1$,
(c) $\beta=0.2$, (d) $\beta=0.5$, (e) $\beta=1$, and (f) $\beta=5$.
The lower and the upper boundaries of the linear phonon band are
shown by the vertical dashed lines and indicated as $1$ and $2$,
respectively.}\vspace{-.3cm}
\end{centering}
\end{figure}
%
\subsection{DBs properties}
Next, we attempt to relate the observed $\beta$-dependent heat
transport behavior to the $\beta$ dependence of DBs properties.
For this aim, we will study DBs properties at both
finite and zero temperatures.
\subsubsection{Evidence for DBs in thermalized chains for $\beta \ge 0.5$}
We first thermalize the focused systems of size $L=200$ (for facilitating the calculation) to the desired temperature $T=0.5$ with the Langevin heat baths~\cite{Lepri_Report,Dhar_Report}. Then, the thermostats are turned off and the power spectrum $P_0(\omega)$ of these thermal oscillations are calculated. In Fig.~\ref{fig6} we present the result of $P_0(\omega)$ versus $\omega$ for several $\beta$ values. As can be seen, for $\beta \ge 0.5$ a notable blueshift of $P_0(\omega)$ towards the direction of higher frequencies is observed [see Fig.~\ref{fig6}(d)]. This is the signature of appearance of the hard-type anharmonicity DBs with frequencies above the linear phonon band, i.e., in the region marked by 2 in Fig.~\ref{fig6}. The appearance of the blueshift of $P_0(\omega)$ for $\beta\ge 0.5$ coincides with the result of Fig.~\ref{fig4} where  the transition from ballistic to normal heat transport takes place at about $\beta = 0.1$-$0.4$. Depletion of $P_0(\omega)$ in the long-wavelength region at large $\beta$ [see Fig.~\ref{fig6}(d)-(f)] can also be explained by the appearance of DBs. Indeed, DBs act as the pins cutting the chain into the parts and reducing the room available for the long-wavelength phonons.

\subsubsection{Application of absorbing boundary conditions}
There exists a numerical method which directly demonstrates the existence of DBs in thermalized lattices~\cite{DBs-1,DBs-2}. To use this approach, again a system of size $L=200$ is thermalized to $T=0.5$. Then, the thermostat is turned off, but now the absorbing boundary conditions are imposed for a long enough time. With such strategy, all the mobile excitations will be absorbed by the boundaries, while the immobile DBs can show up if they exist. Finally, one then can calculate the power spectrum $P(\omega)$ of these residual  oscillations to identify the corresponding immobile DBs frequencies.%
\begin{figure}
\begin{centering}
\vspace{-.6cm} \includegraphics[width=8.8cm]{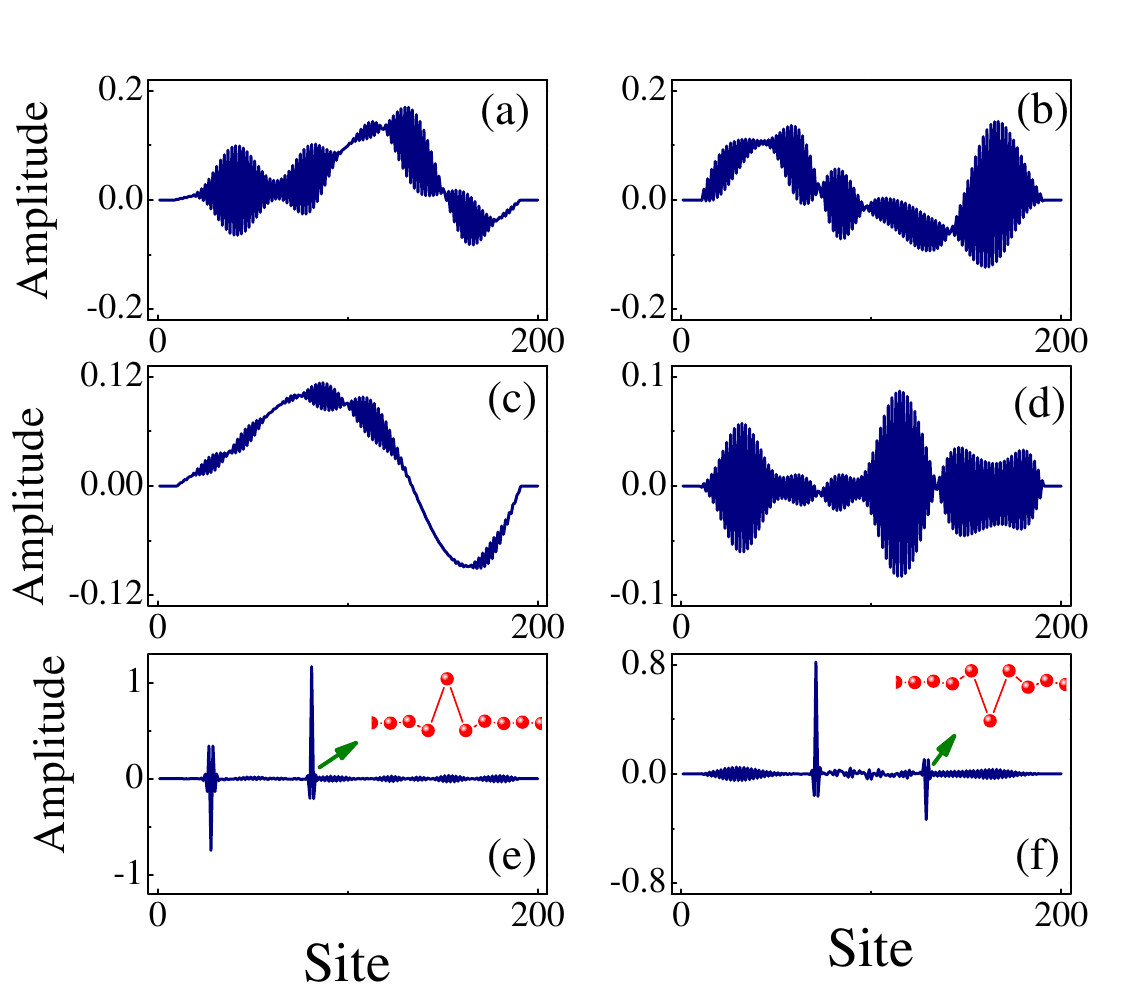} \vspace{-.6cm}
\caption{\label{fig5} Several typical snapshots of the residual
thermal fluctuations after a long time's absorption for three
$\beta$ values, where (a)-(b) for $\beta=0.01$, (c)-(d) for
$\beta=0.2$, and (e)-(f) for $\beta=5$. The insets in
(e) and (f) are used for indicating the Sievers-Takeno DB modes.
}\vspace{-.3cm}
\end{centering}
\end{figure}
%

In Fig.~\ref{fig5} we show several snapshots of the residual thermal
fluctuations after a long time's absorption for three typical
$\beta$ values. For each $\beta$, two typical results from two
different initial equilibrium states of the same focused
temperature are compared. Interestingly, one
can see a variation of such snapshots with $\beta$ as well, i.e.,
for relatively large nonlinearity, we always have the chance to
observe several highly localized modes [see Fig.~\ref{fig5}(e)-(f)].
These modes can be characterized as immobile, stationary DBs
centered on a particle, i.e., the so-called Sievers-Takeno (ST)
modes~\cite{STmode}. For relatively small nonlinearity, the residual
thermal fluctuations are the mixture of phonons with $q$ close to
zero and to the zone boundary ($q=\pm \pi$) [see
Fig.~\ref{fig5}(a)-(d)]. This is expected since such phonons have
vanishing group velocity and they remain in the system after long
exposure to the absorbing boundary conditions.
%
\begin{figure}
\begin{centering}
\vspace{-.6cm} \includegraphics[width=8.8cm]{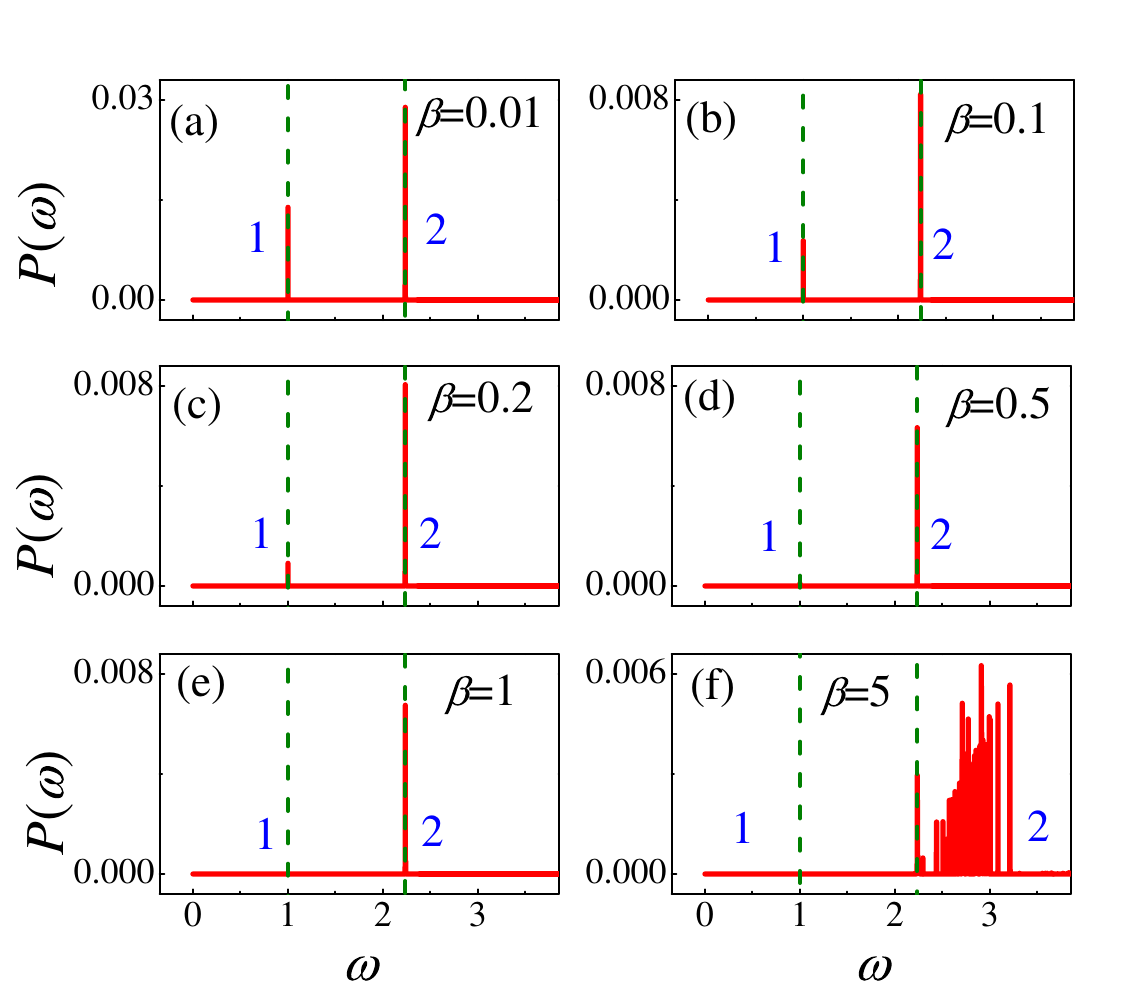} \vspace{-.6cm}
\caption{\label{fig7} The power spectrum $P(\omega)$ of the residual
thermal fluctuations after a long time's absorption: (a)
$\beta=0.01$, (b) $\beta=0.1$, (c) $\beta=0.2$, (d) $\beta=0.5$, (e)
$\beta=1$, and (f) $\beta=5$, where the vertical dashed lines $1$
and $2$ shown the lower and upper edges of the linear phonon band,
respectively. }\vspace{-.3cm}
\end{centering}
\end{figure}
%

We then study the power spectrum $P(\omega)$ of these residual
thermal fluctuations in order to get the information of their
frequencies. The result of $P(\omega)$ versus $\omega$ is shown in
Fig.~\ref{fig7}. Here, similar to Fig.~\ref{fig6}, different values
of $\beta$ are indicated in each panel. From Fig.~\ref{fig7}, first
at a relatively small $\beta$ ($\beta=0.01$), both the peaks
corresponding to the phonons with frequencies close to $q=0$
(denoted by type 1) and to the zone boundary (type 2) can be clearly
identified [see Fig.~\ref{fig7}(a)]. This may explain why the
U-shape can be seen in Fig.~\ref{fig2}(a). Then, with the slight
increase of $\beta$ ($\beta=0.1$ and $\beta=0.2$), the type 1
frequency component becomes weaker and weaker but there is almost no
change for the type 2 frequency component [see
Fig.~\ref{fig7}(b)-(c)], which seems related to the quick damping of
the front peaks of $\rho_Q(m,t)$ shown in Fig.~\ref{fig2}(b)-(c). In
Fig.~\ref{fig7}(d) and (e), the peak corresponding to phonons with
longest wavelengths becomes very weak but the peak for the shortest
wavelengths remains, which might indicate the broader Gaussian peak
of $\rho_Q(m,t)$ as shown in Fig.~\ref{fig2}(d) and (e). Finally,
for the case of strong nonlinearity ($\beta=5$) [see
Fig.~\ref{fig7}(f)], one can clearly see the excitations with
frequencies above the linear phonon band indicating the presence of
immobile DBs in the system in line with Fig.~\ref{fig5}(e)-(f). This
immobile DBs seem to have the effects to make the Gaussian peak
$\rho_Q(m,t)$ narrower [see Fig.~\ref{fig2}(f)]. All of these
evidences clearly show the strong correlation between
$\beta$-dependent DBs properties and heat transport, suggesting that
in this particular system with nonconserved momentum, exploring DBs
properties could be very helpful for understanding heat transport.
That may be also why, previously, researchers often used the role of
DBs as a major phonons scattering mechanism to understand the
diffusive heat conduction in the momentum-nonconserving
systems~\cite{DBs-1,DBs-2}.

Given the strong correlation between the results of Figs.~\ref{fig2}
and~\ref{fig7}, now it is reasonable to assume that DBs are
responsible for the transition to the normal heat transport at large
$\beta$, since in Fig.~\ref{fig7} the role of phonon-phonon
interaction has almost been ruled out, in view of applying the
absorbing boundary conditions. However, the absence of DBs
frequencies in Fig.~\ref{fig7}(d) and (e) should be considered
puzzling, because at $\beta=0.5$ and $\beta=1$, according to
Fig.~\ref{fig4}, we already have normal heat transport obeying
Fourier's law. Thus, in the following we need to find an explanation
why DBs cannot be seen in the chain after long action of absorbing
boundary conditions for $\beta=1$ and why they are still presented
for $\beta=5$. Our viewpoint about this is that it may be related to
the DBs mobility, i.e., at the focused temperature, when $\beta=1$,
the mobile DBs are excited, whereas for $\beta=5$, the immobile DBs
dominate. They have different properties, i.e., the mobile DBs can
be scattering with the lowest-frequency phonons, while after this
scattering has been almost finished, the immobile DBs emerge and
their main role is to localize energy and heat. So, when the
absorbing boundary conditions are imposed, eventually only the
immobile DBs can be identified. As we will show this seems to be
verified in the following.
\subsubsection{DBs at zero temperature: Standing DBs}
With the above puzzle in mind, i.e., the absence of DB frequencies
in Fig.~\ref{fig7}(d)(e), we next explore DBs properties at zero
temperature (for facilitating the analysis). As mentioned, our main
goal is to explain why standing DBs remain in the chain after
applying absorbing boundary conditions in the case of $\beta=5$,
while they cannot be seen for $\beta=1$, cf., (e) and (f) in
Fig.~\ref{fig7}.
\begin{table}[!hbp]
\begin{centering}
\begin{tabular}{ p{1.5cm}||p{1.5cm}|p{1.5cm}|p{1.5cm}}
 \multicolumn{4}{c}{$\beta=1$} \\
 \hline
 $\,A_{\rm DB}$ & $\,\,\,\theta$ & $\,\omega_{\rm DB}$ & $\,E_{\rm DB}$ \\
 \hline
 $\,$0.25  & $\,$0.154  & $\,$2.240 & $\,$  2.028  \\
 $\,$0.5   & $\,$0.309   & $\,$2.257 & $\,$  4.091  \\
 $\,$0.75  & $\,$0.474   & $\,$2.282 & $\,$  6.069  \\
 $\,$1.0   & $\,$0.639   & $\,$2.319 & $\,$  8.147  \\
 $\,$1.25  & $\,$0.871   & $\,$2.354 & $\,$  9.487  \\
 $\,$1.5   & $\,$1.135   & $\,$2.410 & $\,$  10.744 \\
 \hline
 \multicolumn{4}{c}{$\,$} \\
 \multicolumn{4}{c}{$\beta=5$} \\
 \hline
 $\,A_{\rm DB}$ & $\,\,\,\theta$ & $\,\omega_{\rm DB}$ & $\,E_{\rm DB}$ \\
 \hline
 $\,$0.25  & $\,$0.349  & $\,$2.262 & $\,$  0.909  \\
 $\,$0.5   & $\,$0.748   & $\,$2.332 & $\,$  1.751  \\
 $\,$0.75  & $\,$1.360   & $\,$2.451 & $\,$  2.318  \\
 $\,$1.0   & $\,$1.960   & $\,$2.687 & $\,$  3.570  \\
 $\,$1.25  & $\,$2.441   & $\,$3.009 & $\,$  6.091  \\
 $\,$1.5   & $\,$2.814   & $\,$3.380 & $\,$  10.344 \\
 \hline
\end{tabular}
\caption{\label{T1} The standing DB's relevant parameters.}
\end{centering}
\end{table}

We first focus on the properties of standing DBs and in the
following we will try to boost them. An efficient way to excite such
type of standing DBs is the use of the following initial conditions
\begin{equation} \label{Qs}
\phi_k(0)=\frac{(-1)^k A_{\rm DB}}{\cosh[\theta (k-x_0)]}, \quad
\frac{\rm{d} {\phi}_{\mit{k}}(0)} {\rm{d} \mit{t}}=0.
\end{equation}
Here, $A_{\rm DB}$ and $\theta$ are the DB's amplitude and inverse
width, respectively. DB's initial position is located at $x_0$ and
for $x_0=(L-1)/2$ [$x_0=(L+1)/2$] the DB is centered on a particle
(at the center of a bond). We always took $ x_0=(L-1)/2$ to obtain
the ST mode observed in our simulations. For the chosen $A_{\rm
DB}$, we find $\theta$ by using the try and error
method~\cite{JETPL} minimising the oscillations of the DB's
amplitude in simulations. After $\theta$ has been found, we then
calculate DB's frequency, $\omega_{\rm DB}$, and its total (kinetic
plus potential) energy, $E_{\rm DB}$. These results are presented in
Table~\ref{T1} for a set of DB's amplitudes for the focused two
values of $\beta=1$ and $\beta=5$.

From Table~\ref{T1} it can be seen that, with the increase of DB's
amplitude, the degree of its spatial localization, characterized by
$\theta$, increases. The same is true for both the DB's frequency
and energy. In addition, DBs of the same amplitude in the chain with
higher strength of the nonlinearity have higher spatial localization
and higher frequency. However, this is not the case for the DB's
energy. Within the studied range of amplitudes, the energy of DBs
with same amplitude in the chain with weaker nonlinearity is larger,
which shows a big difference between the cases of $\beta=1$ and
$\beta=5$. This is consistent with the results of Fig.~\ref{fig6}(e)
and (f), where the spectrum energies in the case of $\beta=1$ are
obviously larger than the counterparts of $\beta=5$. Regardless of
this difference, in short, all of the data indicate the
concentration of energies played by DBs as shown in
Fig.~\ref{fig2}(f) for the relatively large strengths of the
nonlinearity.
\subsubsection{DBs at zero temperature: Moving DBs}
%
\begin{figure}
\begin{centering}
\vspace{-.6cm} \includegraphics[width=8.8cm]{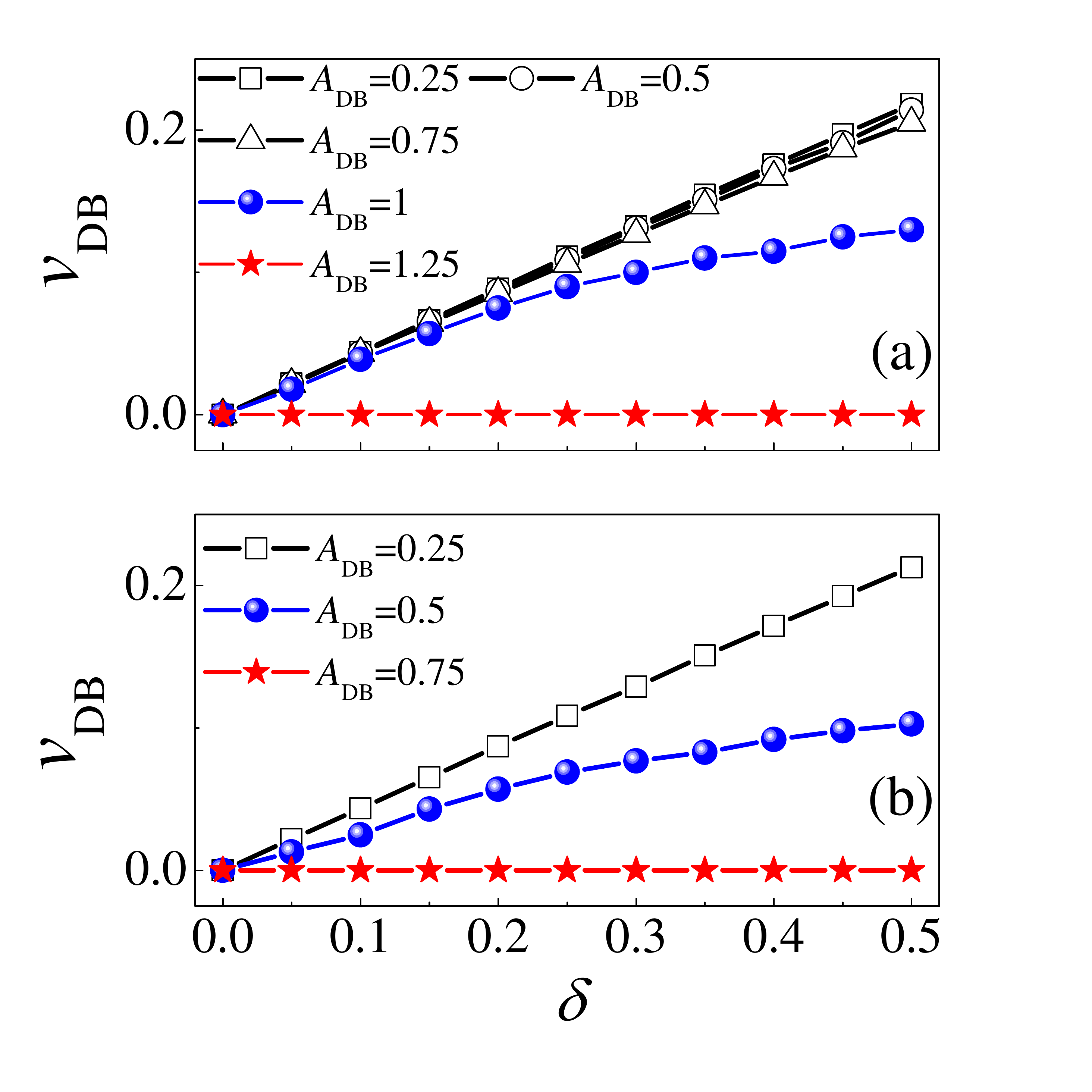} \vspace{-.6cm}
\caption{\label{fig8} Velocity $v_{\rm DB}$ of DBs excited with the
help of the ansatz Eq.~\eqref{Qm} as a function of $\delta$: (a)
$\beta=1$ and (b) $\beta=5$. DB's amplitude is indicated in the
legends. Other DB's parameters are taken from Table~\ref{T1}.
}\vspace{-.3cm}
\end{centering}
\end{figure}
Moving DBs were excited
with the use of the following physically motivated
ansatz~\cite{JETPL}:
\begin{equation} \label{Qm}
\phi_k(t)=\frac{(-1)^k A_{\rm DB}\cos[\omega_{\rm DB} t+\delta(k-x_0)]}{\cosh[\theta(k-x_0)]}.
\end{equation}
Here, $\delta$ is a free parameter which defines the oscillation
phase difference for neighboring particles and also characterizes
DB's velocity, $v_{\rm DB}$, in case when it is mobile. For example, for $\delta=0$,
DB's velocity is zero.

We use Eq.~\eqref{Qm} with different values of $\delta$ for setting
the initial conditions taking other DB's parameters from
Table~\ref{T1}. Velocity of the resulting DB is measured and
presented as a function of $\delta$ in Fig.~\ref{fig8}(a) and (b)
for the chains with $\beta=1$ and $\beta=5$, respectively. Different
lines show the results for different DB's amplitudes $A_{\rm DB}$,
as indicated in the legends. It can be seen that DBs with relatively
small amplitudes have velocities nearly proportional to $\delta$
within the range of $ |\delta| \le 0.5$ considered. For $\beta=1$,
this is true for $A_{\rm DB}\le 0.75$, while for $\beta=5$ it is
observed for $A_{\rm DB}\le 0.25$. Such DBs are highly mobile. We
have checked that they move through entire computational cell of
3000 sites with nearly constant velocity and practically radiating
no energy. However, for the cases of, e.g., $A_{\rm DB}=1$
($\beta=1$) and $A_{\rm DB}=0.5$ ($\beta=5$), DB's velocity
saturates with the increase of $\delta$, it radiates small-amplitude
waves, and its velocity gradually decreases. For this reason, we can
only measure the DB's velocity up to $t=400$. Finally, DBs with even
higher amplitudes are trapped by the lattice and no longer move for
any value of $\delta$. This is observed for $A_{\rm DB}\ge 1.25$
($\beta=1$) and $A_{\rm DB}\ge 0.75$ ($\beta=5$).

Combining the results presented in Fig.~\ref{fig8} and the DB's
parameters shown in Table~\ref{T1}, now one may recognize that the
immobility of DBs takes place for about $\theta \ge 0.7$. Indeed,
Table~\ref{T1} indicates that, for $A_{\rm DB}=1$ ($\beta=1$), one
has $\theta=0.639$; and for $A_{\rm DB}=0.5$ ($\beta=5$),
$\theta=0.748$. Both cases are close to the transition of DBs from
mobile to immobile as suggested by Fig.~\ref{fig8}.

It is also very interesting to note that at the transition point of
$\theta \simeq 0.7$, the DB's energy for $\beta=1$ ($E_{\rm
DB}=8.147$) is much higher than that in the case of $\beta=5$
($E_{\rm DB}=1.751$). This may correspond to the fact that the
probability of excitation of immobile DBs for $\beta=1$ is
relatively small compared to the case of $\beta=5$, consistent with
the results shown in Fig.~\ref{fig7}(d)-(f).

\subsubsection{Acceleration of DBs by phonons}
\begin{figure}
\begin{centering}
\vspace{-.3cm} \includegraphics[width=8.8cm]{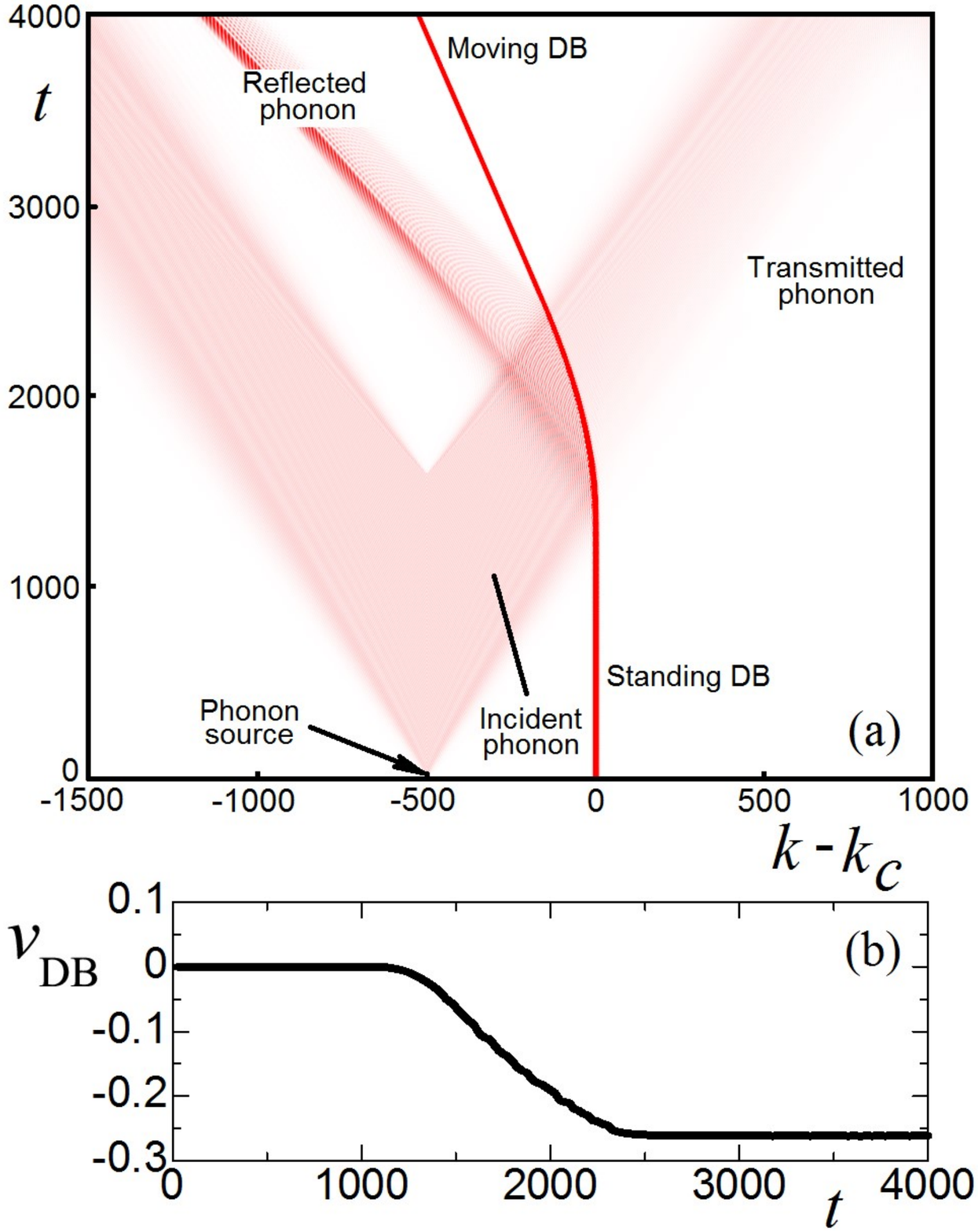} \vspace{-.6cm}
\caption{\label{fig9} (a) Counter plot showing the evolution of
total (kinetic plus potential) energies of particles with $t$ in the
chain with $\beta=1$. Initially ($t=0$), in the center $k_c$, a
standing DB is excited by using ansatz~\eqref{Qm} with $A_{\rm
DB}=0.5$, $\delta=0$, $\omega_{\rm DB}=2.257$ and $\theta=0.309$,
and an ac driving phonon source at $k^{*}=k_c-500$ is operated until
$t=1500$. One can see that the DB is accelerated towards the phonon
source during the scattering with the phonon wave packet. (b) Time
evolution of DB's velocity during this process. }\vspace{-.6cm}
\end{centering}
\end{figure}
Now we understand that there are two different types of DBs existing
in the system characterized by their different mobility induced by
different $\beta$ values together with the different $A_{\rm DB}$.
With this understanding, we next demonstrate that the mobile DBs can
be scattered by the small-amplitude phonons. To see this, we first
excite a DB in the center [$k_c=(L-1)/2$] of the chain. As an
example, we choose the case of $\beta=1$ and set $A_{\rm DB}=0.5$
and $\delta=0$, so $\theta=0.309$ according to Table~\ref{T1}. Thus,
a mobile DB with zero initial velocity according to Fig.~\ref{fig8}
is excited. At the same time, on the left-hand side of the DB, a
particle with the index number $k^{*}$ is forced to move according
to the following ac driving:
\begin{equation} \label{ACdriving}
\phi_{k^*}(t)=A_{\rm p}\sin(\omega_{\rm p} t),
\end{equation}
where $A_{\rm p}$ and $\omega_{\rm p}$ are the driving amplitude and
frequency, respectively. We use a relatively small $A_{\rm p}$ and
set $\omega_{\rm p}$ within the phonon band, so that a phonon-like
wave packet with the given frequency is excited and interacts with
the DB. In order to provide the details of such scattering process,
we then record each particle's energy for a long time up to
$t=4000$. A typical result with the parameters $A_{\rm p}=0.075$,
$\omega_{\rm p}=1.1$ and $k^*=k_c-500$ is shown in
Fig.~\ref{fig9}(a). Under this setup, we perform the ac driving for
a time lag up to $t=1500$ and then it is stopped.

From Fig.~\ref{fig9}(a), one can see that the incident phonon
wave packet is partially reflected and partially
transmits, which is reasonable. While for the DB, the result is very
interesting. Due to the scattering by the phonon wave packet, the
standing mobile DB is accelerated towards the phonon source. This
indicates that the phonon-DB interaction can cause the mobile DB from
standing to moving, and in turn provides a scattering mechanism for
phonons. For more details, we also detect the time evolution of the
DB's velocity, which is shown in Fig.~\ref{fig9}(b). This result
tells us that, a standing DB after interaction with the phonon wave
packet moves with a constant nonzero velocity.

\subsubsection{Summary of DBs properties in relation to heat transport}
Now we summarize our main results on DB properties and try to relate
them to heat transport. In the focused $\phi^{4}$ model with
hard-type anharmoncity ($\beta>0$), only DBs with frequencies above
the linear phonon band can exist. While including on-site potential
makes these DBs properties quite peculiar, which seem distinct from
those shown in the momentum-conserving systems. Our numerical
analysis suggests that, (i) DBs with high degree of spatial
localization ($\theta \ge 0.7$) are immobile, while less localized
DBs ($\theta \leq 0.7$) are mobile. So, $\theta \approx 0.7$ is a
threshold value at which DBs change from mobile to immobile. This is
indeed true for the two typical cases we considered, i.e., $A_{\rm
DB}\ge 1.0$ for $\beta=1$ and $A_{\rm DB}\ge 0.5$ for $\beta=5$.
(ii) At the transition point of $\theta \approx 0.7$, the single
DB's energy $E_{\rm DB}$ in the chain with $\beta=1$ is almost five
times larger than that of $\beta=5$. Then, according to the
Arrhenius law~\cite{ArrLaw}, the probability of DB's excitation by
thermal fluctuations at a finite temperature $T$ is proportional to
${e}^{-E_{\rm DB}/(k_B T)}$, where $k_B$ is the Boltzmann constant.
This explains why for $\beta=1$ the probability of excitation of
immobile DBs is orders of magnitude smaller than in the case
$\beta=5$. So, for $\beta=1$ DBs are mainly mobile and it is
reasonable to see that they disappear after application of the
absorbing boundary conditions for a long time, as evidenced in
Fig.~\ref{fig7}(e); whereas in the case of $\beta=5$ we have a large
chance to excite the immobile DBs, so they show up in
Fig.~\ref{fig5}(e)-(f) and Fig.~\ref{fig7}(f) after exposure to the
absorbing boundary conditions for a long time. (iii) DBs can be
scattered with phonons. In particular, the mobile DBs can be
accelerated by the phonons as exemplified by Fig.~\ref{fig9}. This
in turn causes a scattering mechanism for phonons. So, when this
scattering becomes dominant, the normal heat transport can be
observed. This is indeed the case for $0.4 \leq \beta \leq 1$ as
shown in Fig.~\ref{fig2}(d) and (e). For even higher strength of the
nonlinearity ($\beta=5$ for example), the immobile DBs are readily
excited and normal heat transport is now characterized by a narrower
Gaussian peak, as shown in Fig.~\ref{fig2}(f). This may suggest that
the immobile DBs can localize the energy and heat.
\section{Discussion}
\label{SecDiscussion}
Finally, we discuss why the above pictures are related to the
nonconservation of momentum and what happens in the range of $\beta \leq 0.4$. For this aim we employ the momentum
correlation function $\rho_p(m,t)$ to detect the relevant
information. Similarly to $\rho_Q(m,t)$, $\rho_p(m,t)$ is defined by
\begin{equation} \label{PP}
\rho_{p} (m,t)=\frac{\langle \Delta p_{j}(t) \Delta p_{i}(0)
\rangle}{\langle \Delta p_{i}(0) \Delta p_{i}(0) \rangle}.
\end{equation}
Here $\Delta p_{i}(t) \equiv p_i(t) - \langle p_i \rangle$ denotes
the momentum fluctuation at a bin $i$ and time $t$. Simulation of
this correlation function is also similar to that of $\rho_Q(m,t)$.

$\rho_{p} (m,t)$ has
been verified to be very useful in understanding anomalous thermal
transport. For example, from the perspective of hydrodynamics
theory~\cite{Chen2013,Spohn2014}, $\rho_{p} (m,t)$ has been
conjectured to represent the sound modes' correlation. A diffusive
behavior of $\rho_{p} (m,t)$ has been observed in the couple rotator
systems and related to the observed normal heat
transport~\cite{YunyunLi2015}. Such non-ballistic diffusive momentum
spread appears to be the origin of the recovery of Fourier's law in
a special system with a double-well interparticle
potential~\cite{Xiong2016-1,Xiong2016-2}. A more recent work based
on an effective linear stochastic structure theory~\cite{DCai2016}
has indicated that the scaling behavior of $\rho_{p} (m,t)$ can
enable us to explore the sound damping information, and thus is
related to anomalous heat transport. However, all of the above results are the understanding for momentum-conserving systems,
here, however, we will use $\rho_{p}
(m,t)$ to understand the origin of normal heat transport in the
momentum-nonconserving systems.
\begin{figure}
\begin{centering}
\vspace{-.6cm} \includegraphics[width=8.8cm]{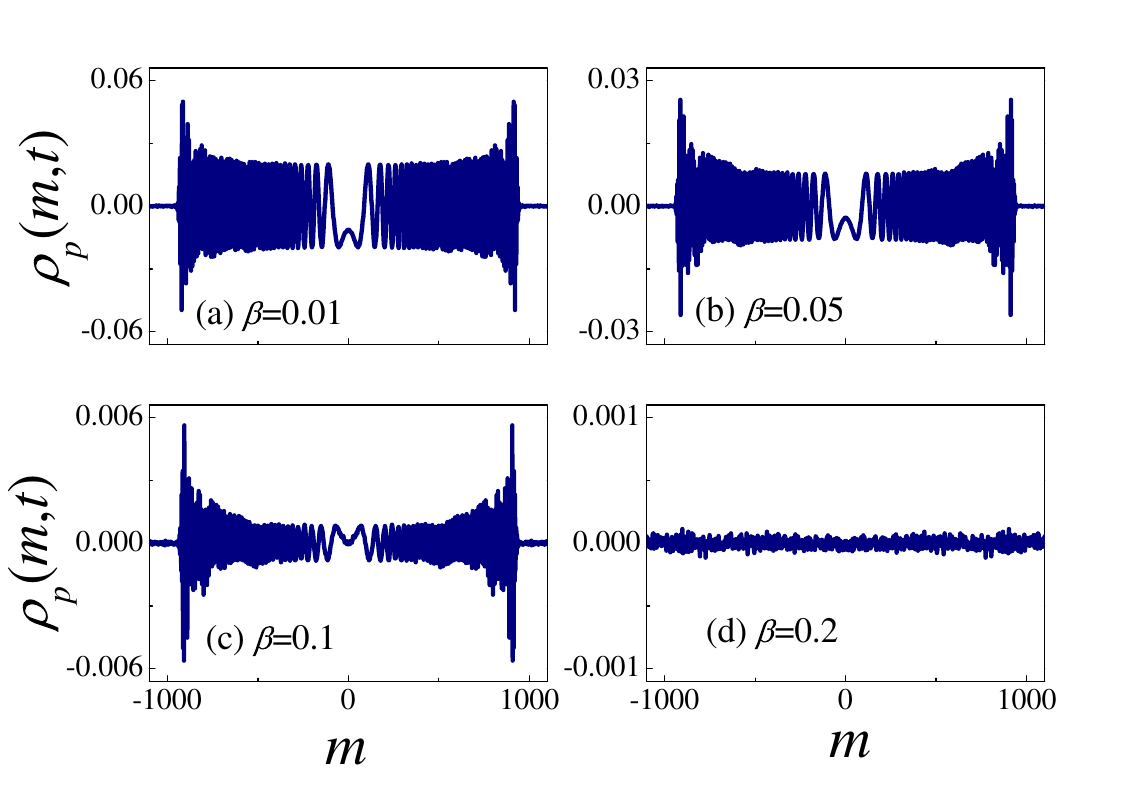} \vspace{-.6cm}
\caption{\label{fig10} The momentum correlation function
$\rho_p(m,t)$ for several small $\beta$ values for a long time
$t=1500$: (a) $\beta=0.01$, (b) $\beta=0.05$, (c) $\beta=0.1$, and
(d) $\beta=0.2$. }\vspace{-.3cm}
\end{centering}
\end{figure}

Figure~\ref{fig10} presents the result of $\rho_p(m,t)$ for several
small $\beta$ values. It is easy to find that above $\beta=0.2$, all
the correlation information of $\rho_p(m,t)$ disappear. This is so
because, in this range the mobile DBs become dominated. Then, the
phonons with low frequencies, i.e., the sound modes, can be
completely damped due to the scattering with DBs. This is another
type of signature of normal heat transport in momentum-nonconserving
systems induced by the on-site potentials, different from the
diffusive behavior of $\rho_p(m,t)$ shown in some
momentum-conserving
systems~\cite{YunyunLi2015,Xiong2016-1,Xiong2016-2}. In the context
of hydrodynamics, it means a complete damping of the sound modes'
correlation, similarly to the results of
~\cite{YunyunLi2015,Xiong2016-1,Xiong2016-2} but obviously here it
is in a distinctive way.

Now, inspired by this macroscopic evidence of $\rho_p(m,t)$, we may
conjecture that, with slight nonlinearity ($0.01 \leq \beta \leq
0.4$), there are also mobile DBs excited in the system. While these
mobile DBs are not dominated, so, the scattering of phonons are not
strong enough to cause normal heat transport. Therefore, a gradual
damping of the momentum correlation information as shown in
Fig.~\ref{fig10} can be observed. Finally, as to the full underlying
picture in the whole ranges of $0.01 \leq \beta \leq 5$, we would
like to suggest such an understanding: To induce normal heat
transport in the momentum-nonconserving $\phi^{4}$ system, one
should resort to the nonlinearity. This nonlinearity together with
the on-site potentials can cause a phonons scattering mechanism
induced by the mobile DBs with frequency components slightly above
the linear phonons band. For the very large nonlinearity, after the
scattering process has been almost completed, another type of
immobile DBs with frequency components greatly upper the linear
phonon band now becomes dominated. This kind of immobile DBs can
localize the energy and heat~\cite{Xiong-DBs}, which may make the
system like an insulator with a very low thermal conductivity. So, a
narrow Gaussian heat transport density can be observed in
Fig.~\ref{fig2}(f).
\section{Conclusions} \label{SecConclusions}
In summary, we have succeeded in understanding the transition
process to normal heat transport in the momentum-nonconserving
$\phi^{4}$ system when the strength of the hard-type nonlinearity
$\beta$ is increased. To do this, we have examined both the
macroscopic heat transport property and the microscopic details of
DBs dynamics for different $\beta$ values. In general, as $\beta$
increases, we have found that, the heat transport can undergo a
crossover from ballistic to normal, with some rich details, i.e.,
first, the ballistic moving peaks of the heat perturbations
correlation are damped, and then its central part becomes more and
more concentrated. Interestingly, such rich details seem to be
strongly related to the microscopic DBs properties, i.e., the
nonlinearity together with the on-site potential makes the system to
excite both mobile and immobile DBs, respectively, at the relatively
weak and strong nonlinearity. These DBs have different effects on
the heat transport property. On one hand, the mobile DBs induce a
scattering mechanism for phonons which results in the damping of the
sound modes' correlation. Eventually, when this effect of scattering
becomes dominated, a normal thermal transport obeying the Fourier's
law characterized by a Gaussian-type heat transport density can be
seen. On the other hand, for the relatively large nonlinearity, the
main excitations are instead the immobile DBs. This kind of DBs has
the effects of localizing energy and heat, making the chain like an
insulator, and finally reducing the total thermal conductivity.
Therefore, in this range even the Fourier's law is recovered as
well, the Gaussian-type heat spreading density now becomes narrower
and narrower. Such an understanding apparently provides more
detailed evidence for relating the macroscopic heat transport
properties and the underlying microscopic dynamics in the particular
momentum-nonconserving $\phi^{4}$ system.

In spite of above achievement, we should finally note that, to
induce normal heat transport, generally one should rely on both the
effects of nonlinearity and on-site potential, while all of the
above understandings from DBs properties are just based on the
quartic nonlinear on-site potential considered here. Although this
is the mostly popular model used for demonstration the role of
nonconserved momentum~\cite{Momentum-2}, we are still not sure that
whether the proposed picture/explanation has its generality and can
be extended to general models. Hence, in the next step we are
attempted to study the case of on-site potential also including
cubic anharmonicity. In addition, as to the relevant underlying
picture, we here only provide strong evidences that, at zero
temperature the mobile and immobile DBs can exist in the system, and
their contributions to heat transport could be different. A further
more complete picture about the phonon-DBs interaction at
thermalized equilibrium state is still lacking. For this, a recent
fractal L\'{e}vy heat transport picture as suggested in the
nanoparticle embedded semiconductor
alloys~\cite{Ver3,Upadhyaya,Ver1,Ver2} would be insightful.
\begin{acknowledgments}
D.X. was supported by the National Natural Science Foundation of
China (Grant No. 11575046); the Natural Science Foundation of Fujian
Province, China (Grant No. 2017J06002); the Training Plan Fund for
Distinguished Young Researchers from Department of Education, Fujian
Province, China, and the Qishan Scholar Research Fund of Fuzhou
University, China. Stay of D.S. at IMSP RAS was partly supported by the Russian
Science Foundation, grant No. 14-13-00982. S.V.D. was supported by the grant of the Russian
Science Foundation (No. 16-12-10175).
\end{acknowledgments}

\end{document}